\documentclass[%
preprint,
amsmath,
amssymb,
]{revtex4-1}

\usepackage{graphicx}
\usepackage{dcolumn}
\usepackage{bm}

\begin{document}

\preprint{Annalen Der Physik}

\title{Pair Events in Superluminal Optics}

\author{Robert J. Nemiroff}
\affiliation{Department of Physics, Michigan Technological University, Houghton, MI, USA.}

\date{\today}

\begin{abstract}
When an object moves faster than emissions it creates, it may appear at two positions simultaneously. The appearance or disappearance of this bifurcation is referred to as a pair event. Inherently convolved with superluminal motion, pair events have no subluminal counterparts. Common examples of superluminal motions that exhibit pair events include Cherenkov radiation, sonic booms, illumination fronts from variable light sources, and rotating beams. The minimally simple case of pair events from a single massive object is explored here: uniform linear motion. A pair event is perceived when the radial component of the object's speed toward the observer drops from superluminal to subluminal. Emission from the pair creation event will reach the observer before emission from either of the two images created. Potentially observable image pair events are described for sonic booms and Cherenkov light. To date, no detection of discrete images following a projectile pair event have ever been reported, and so the pair event nature of sonic booms and Cherenkov radiation, for example, remains unconfirmed. Recent advances in modern technology have made such pair event tracking feasible. If measured, pair events could provide important information about object distance and history. 
\end{abstract}

\pacs{Valid PACS appear here}
\maketitle


\section{Introduction}

Media conduct emissions. Typically, the fastest that any object can travel inside a given medium is a well-defined speed expressed in terms of that medium's stiffness and density. In this work, however, speeds faster than this emission speed will be considered. Specifically, pair events, a phenomenon where a single projectile suddenly appears twice, and the subsequent pair nature of the resulting images, will be analyzed. 

In popular culture, superluminal speeds are confined to the realm of science fiction. However, there are several very real, massive projectiles that are superluminal in the sense that they actually move through a medium faster than emissions. Perhaps the simplest is a relativistic charged particle entering a dense medium, for example water, and causing the emission of Cherenkov radiation \citep{1937PhRv...52..378C}. An analogous example involving sound is a supersonic plane moving through air (see, for example, \citet{1963AIAAJ...1.1327F}). Note that emissions created by the massive object in its own frame, which would be superluminal in the frame of the ambient medium, are not considered here. Rather, the massive object acts as a power source for emissions, for example a pressure wave, that propagates away spherically at $c_n$, the maximum speed in the medium. In this sense, the superluminal object triggers and powers the release of emissions in the ambient medium (see, for example, \citet{1963flp..book.....F} and \citet{1975clel.book.....J}). 

Other types of superluminal motion involve massless phenomena. For example, the shadow of opaque objects moving past light sources are not confined to be subluminal \citep{1995iqm..book.....G}. More generally, all illumination fronts from discrete flashes are necessarily superluminal \citep{2016PhyEd..51d3005N}. An early mention of pair events was in \citet{1971Sci...173..525C} who hypothesized that apparent superluminal motion evident in quasars might derive from them. The apparent "splitting" of superluminal spots from a rotating source was noted in computer animations by \citet{2009PhyEd..44..296B}, while a pair creation event was actually measured in the lab for a plane wave incident on a tilted screen in a bold experiment by \citet{2016SciA....2E1691C}. The creation of pair event echoes from a sweeping beam was discussed in detail in \citet{2015PASA...32....1N}, and from a flash on a linear reflecting medium by \citet{2017arXiv170305811N}.

The purpose of this work is to explore the concept through a simple example of constant linear motion of a massive object, and to show that these events may be detectable in practice. It is hoped that others will detect and find uses for pair events, although more complex scenarios may be needed than the straight-forward examples considered here.

\section{Radial Motion}

First considered will be linear motion of a projectile moving at constant speed $v$ along a line that includes the observer. This motion will here be considered to be along the $x$ axis, with the projectile moving from negative $x$ to positive $x$, and with the observer located at $x=0$. It is straightforward to compute a type of apparent speed $u$ that helps to illuminate the pair event phenomenon. To define this speed, it is helpful to first picture equally spaced light bulbs located along the projectile's path. As the projectile moves along this line, it passes the bulbs. Suppose that every time a bulb is passed, it flashes, and that light from each flash, moving isotropically out from the bulb at a speed of magnitude $c$, eventually reaches the observer. Noting the time the observer sees each flash along with the distance between bulbs allows the observer to compute this apparent speed. 

Mathematically, relative to an observer at the origin, it is straightforward to find that the perceived speed of approach of the projectile would be 
\begin{equation} \label{uapproach}
u_{approach}  = \frac{v}{(1 - v/c)} , 
\end{equation}
from the $-x$ direction as seen by the observer, while the perceived speed of retreat would be 
\begin{equation} \label{uretreat}
u_{retreat} = \frac{v}{(1 + v/c)}, 
\end{equation}
which would be observed toward the $+x$ direction.

To better understand what happens at $v > c$, it is of interest to first consider slower object motions. For non-relativistic motion, $v << c$, then Eqs. (\ref{uapproach}) and (\ref{uretreat}) show that both $u_{approach}$ and and $u_{retreat}$ are positive and approach $v$, as expected. 

For relativistic but subluminal speeds with $v < c$, Eqs. (\ref{uapproach}) and (\ref{uretreat}) show that $u_{approach}$ is always positive but can exceed $c$ (see, for example, \citet{1977Natur.267..211B}
). Also $u_{approach}$ can be significantly larger than $u_{retreat}$, and the later must be less than $c$. Discerning the directions of approach and retreat of the real object are straightforward, since approach is observed to occur before retreat. 

At the speed of light, When $v = c$, then the observer first sees the projectile only at the exact moment it passes. Because $u_{approach}$ diverges, nothing is ever visible from the direction of approach. The object is only seen to retreat at $u_{retreat} = c/2$. 

When $v > c$, the observer again sees the projectile first only when it passes. What happens next, though, is a pair event, which might be counter-intuitive. Eqs. (\ref{uapproach}) and (\ref{uretreat}) show that two images of the object suddenly appear at the observer and each moves away, simultaneously, in opposite directions. That $u_{approach}  < 0$ indicates that even the "approaching" image is actually seen to move away, toward increasingly negative $x$ values. The observer notes a dilemma if trying to discern from which direction the passing object really approached. However, because $|u_{approach}| > |u_{retreat}|$, the observer can still tell the direction of approach by comparing the magnitudes of the apparent speeds. 

Summarizing conceptually, if an object is moving directly toward an observer in a medium faster than emissions it triggers in that medium, then it will reach the observer before any of these emissions. Subsequently, if the observer is looking in the direction from which the projectile approached, the observer will perceive the projectile to be moving away, and not toward, the observer. This is because radiation emitted increasingly earlier only reaches the observer consecutively later. Moreover, were the same observer to look at the same projectile in the direction that it actually moves away, the observer will see a second image of it -- also moving away. This is because radiation emitted after passing the observer also reaches the observer consecutively later. Therefore, if the observer could look in both directions simultaneously, the observer would never see the superluminal projectile approach, but would suddenly see two images of the object appear and recede, one in each direction. 

\section{Non-Radial Motion}

\subsection{Speed}

\begin{figure} 
     \includegraphics[width=0.8 \columnwidth]{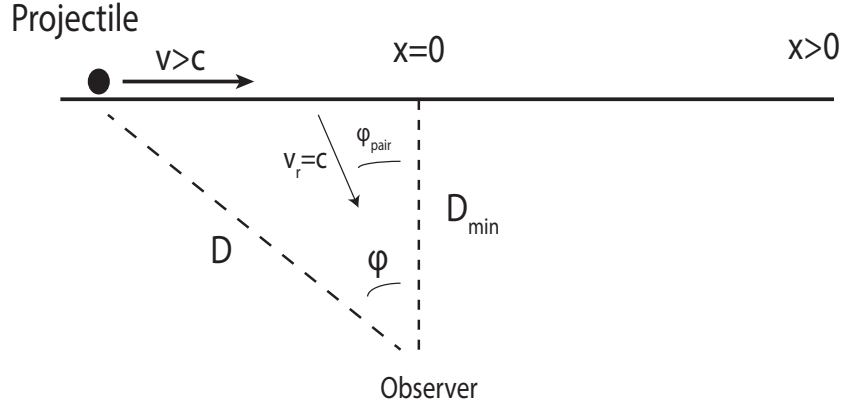}
     \caption{The geometry of the projectile moving linearly and at constant speed $v$ in the view of the observer. The pair event is seen by the observer at $\phi = \phi_{pair}$, when $v_r = c$.}
     \label{PairGeometry}
\end{figure}

If the projectile moves in a line that does {\it not} intersect the observer, the situation is a bit more complex. For completeness, it will be assumed that the projectile starts infinitely far from the observer and moves at a constant speed $v$ in a line that takes it to a minimum distance $D_{min}$ from the observer, after which it moves off again along the same line. In terms of the angular perspective of an observer at the origin, the projectile moves from $\phi = -\pi/2$ to $+\pi/2$ and always has positive speed. As in the radial case, images that move toward lower $\phi$ values will be considered to have negative speed. The geometry is depicted in Figure~\ref{PairGeometry}

When the radial component of the approaching projectile's speed toward the observer is faster than $c$, so that $v_r > c$, the object will always be closer to the observer than its emissions. Therefore, the object will appear to the observer to be moving away. As the projectile nears the observer, $v_r$ will decrease, reaching zero at $D_{min}$. Now when $v_r < c$, emission from the object will precede the object and so the object will appear to the observer to be moving closer. Therefore, the first time this object is seen by the observer is when $v_r = c$. This is referred to as a perceived pair creation event, or just "pair event". The angular location of the pair event will be designated $\phi_{pair}$. After the pair event, the observer sees two images of the object moving away from $\phi_{pair}$. One image first approaches the observer, moves with increasing $\phi$, passing $\phi = 0$ -- the angle corresponding to the distance of closest approach -- and then appears to recede from the observer. A second image starts at $\phi_{pair}$ and moves with decreasing $\phi$, always receding from the observer.

The location of the pair event was derived by \citet{2015PASA...32....1N} to be 
\begin{equation} \label{Eq:PhiPair}
\phi_{pair} = -  \arcsin(c/v) .
\end{equation}
Note that when $v$ diverges, $\phi_{pair}$ approaches zero, the direction toward closest approach. Also, there are no real solutions for $v < c$ since no perceived pair events will appear. 

The apparent speed $u$ of this non-radially moving projectile can be computed in a similar fashion as the radial case. Suppose, again, that there are equally spaced light bulbs located along the projectile's line of motion. Suppose, again, that every time a light bulb is passed, it flashes, and that light from these flashes eventually reaches the observer. The observer will again note the times of receiving flashes, and from knowing the spacing of the light bulbs, can compute the apparent speed $u$ of the speeding object.

Following the analysis of a light beam sweeping across a wall given in \citet{2015PASA...32....1N}, the apparent speed $u$ of a projectile with real constant speed $v$ in linear motion is found to be 
\begin{equation} \label{UNonRadial}
u = \frac{c v}{v \sin \phi + c} .
\end{equation}
Note that Eq. (\ref{UNonRadial}) reduces to Eqs. (\ref{uapproach}) and (\ref{uretreat}) for radial motion: when $\phi = - \pi/2$, on approach and $\phi = +\pi/2$ on retreat. The non-relativistic limit, when $v << c$, yields $u = v$, as expected. At the pair event, $u$ diverges. Note also that pair events can only occur when $v \ge c$, because when $v < c$ the denominator never diverges. 
\begin{figure} 
     \includegraphics[width=0.8 \columnwidth, angle=90]{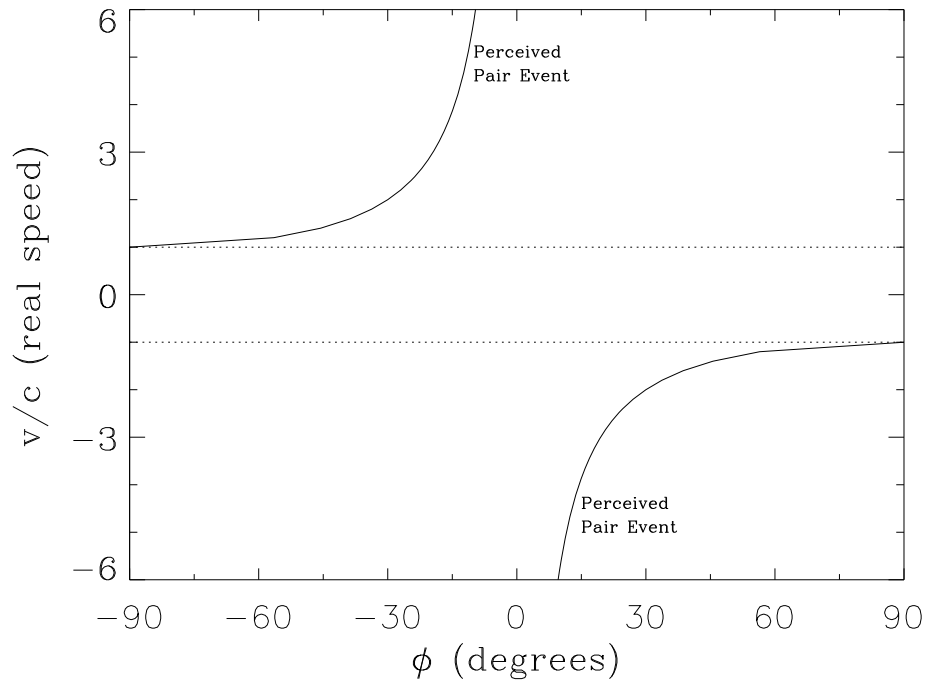}
     \caption{A plot of the real speed $v$ of the projectile verses the viewing angle where a pair creation event is seen to occur by an observer. The angle $\phi = 0$ corresponds to closest approach. The faster the projectile, the closer to $\phi = 0$ the pair event will be observed.}
     \label{fig:phipair}
\end{figure}

Figure~\ref{fig:phipair} shows where the observer would perceive the pair event to occur, given different projectile velocities. For extremely high values of linear speed $v >> c$, pair events occur near $\phi = 0$, whereas for lower values of $v$ just above $c$, pair events occur at lower values of $\phi$. Negative values of $v$ refer to a projectile moving from negative $\phi$ to positive $\phi$, whereas positive values of $v$ refer to a projectile moving from positive to negative $\phi$. No $\phi_{pair}$ will exist for $v < c$.
\begin{figure} 
	\includegraphics[width=0.8 \columnwidth, angle=90]{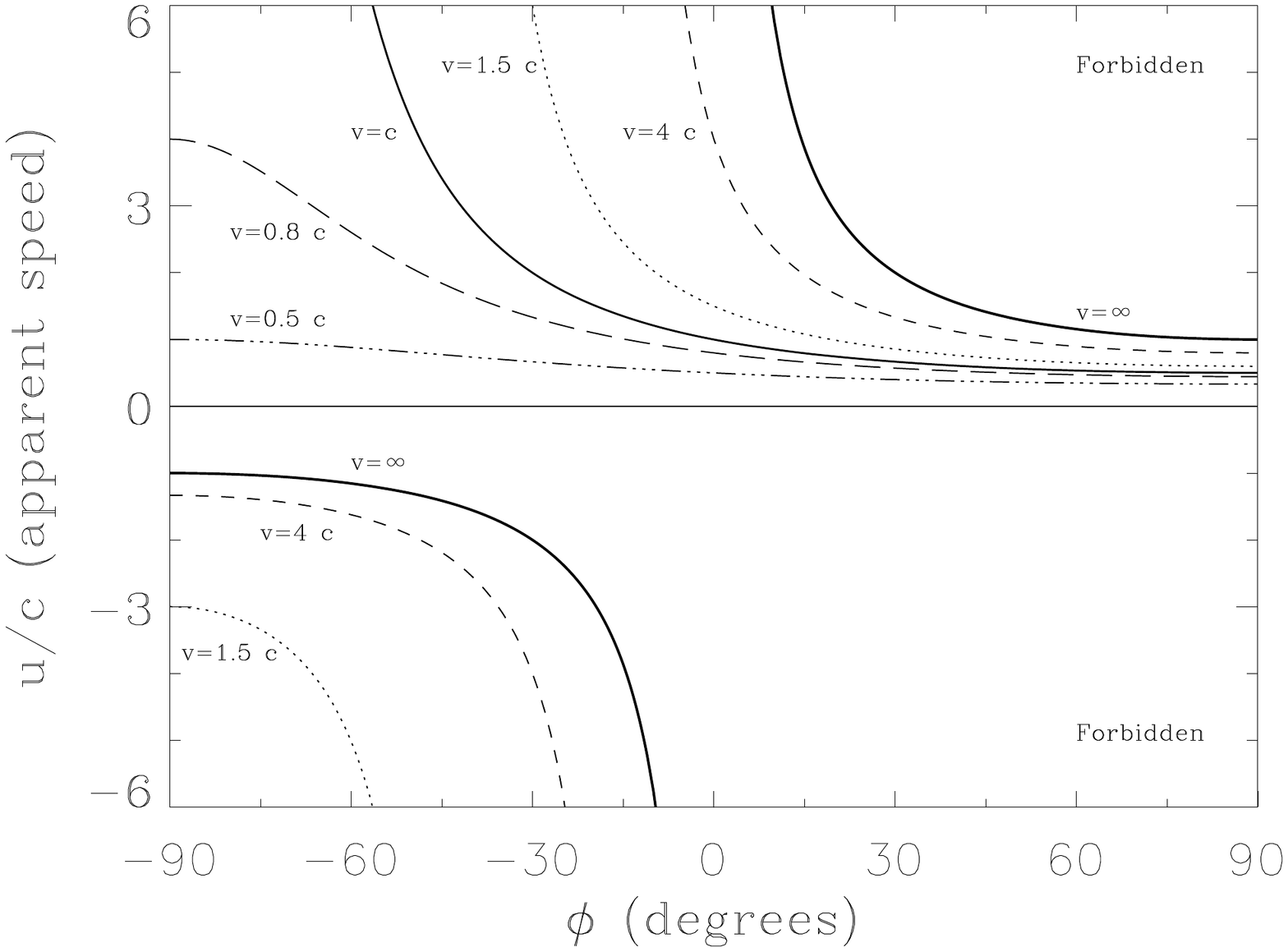}
    \caption{A plot of the apparent speed $u$ of the projectile versus viewing angle $\phi$. The highest observable apparent speeds occur for projectiles moving just faster than $c$ and formally diverge at the pair event.}
    \label{fig:superapparent}
\end{figure}

Figure~\ref{fig:superapparent} depicts the apparent speed $u$ of images as a function of viewing angle $\phi$. The projectile is assumed to be moving at constant speed $v$ from negative $\phi$ to positive $\phi$, with closest approach to the observer at $\phi = 0$. Negative values of $u$ correspond to the observer at the origin observing an image moving from positive $\phi$ to negative $\phi$. For values of $v < c$ only one image exists, whereas for values of $v> c$, two images exist -- after the pair event -- each moving in opposite directions. 

\subsection{Brightness}

As the projectile moves at a constant speed through a uniform medium, emissions it excites in the medium are continually created. Assuming that these emissions radiate away isotropically in the rest frame of the medium, they are analogous to the bulbs lit by the passing projectile discussed above. Just as each light bulb itself emits the same brightness as it is passed, the intrinsic brightness in a medium created by a passing projectile is constant and isotropic. It is only how these emissions appear to the observer that creates the perceived pair event "flash" for fast-enough projectiles. Therefore, the instantaneous brightness of an image to an observer, integrated over a given period of time, will be proportional to the instantaneous transverse apparent speed of an image as perceived by the observer. Since this speed formally diverges at the pair event, as perceived by the observer, the pair event appears to the observer to be, formally, infinitely bright, albeit for only an infinitesimal amount of time. More generally, increasing geometric distance will cause the apparent instantaneous brightness to vary inversely with the square of the distance $D$ to the observer. Neglecting emanations from the superluminal particle itself, the integrated perceived brightness of an image between times $t_1$ and $t_2$ in the directions of their corresponding $\phi$ values follows 
\begin{equation} \label{Eq:b}
b \propto \int_{t_1}^{t_2} \frac{u_t \ dt}{D^2} ,
\end{equation}
where $u_t = u \cos \phi$ is the transverse component of the perceived speed of the image.

\section{Examples}

\subsection{Sonic Booms}

More people have {\it heard} a pair event than seen one. This is because a pair event in sound is referred to as a "sonic boom". When an airplane is moving linearly above the speed of sound in air, it creates pressure waves that move out spherically from its instantaneous location. A supersonic airplane actually creates a double-boom from a pressure "N-wave" \citep{haering2006flight}, with a high pressure region leading the plane, and a low pressure region trailing -- however for simplicity they will be considered a single wave. The superposition of these pressure waves is called the Mach cone, but what is heard by an observer at the leading edge of this cone is a sonic boom. 

A sonic boom results from a pair event. Specifically, air molecules are analogous to the light bulbs mentioned previously. A sonic-boom pair event is the first thing an observer hears, occurring when the airplane's radial speed toward the observer drops from supersonic to subsonic. After the boom, an observer would hear the airplane from two locations simultaneously, as each sound "image" moved away from the sonic-boom pair-event direction. The effect was first noted with respect to computer-modeled wavefronts by \citet{ahrens2008reproduction}. An interesting historical note is that in 1896 Lord Rayleigh, although not mentioning that {\it two} sound images would occur, did note that a supersonic sound could be heard backwards \citep{rayleigh1896theory}. The backwards nature of one of the sound images was highlighted in a "What If" xkdc comic \citep{2013Whatif...March19}. To the best of the author's knowledge, however, the dual nature of sound pair events has never yet been recovered experimentally. A primary reason is because the identification of sonic booms as pair events is relatively unknown. However, a simple verification experiment might confirm this. 

As an example, assume the supersonic airplane moves linearly at a constant Mach 2 through air, passing an observer at a minimum overhead distance of $D_{min} = 10,000$ meters. Although sound speed decreases slightly with altitude, it will be assumed constant at $c = 320$ m sec$^{-1}$. Defining the observer to reside at $\phi = 0$, the sonic boom would be heard coming from the direction $\phi_{pair} = -30^o$. Following Eq. (6) in \citet{2015PASA...32....1N}, one can find the times and positions of each sound image. Specifically, five seconds after the boom was heard, two separate sound images will reach the observer simultaneously from directions of $\phi \sim -54.8^o$ and $\phi \sim 2.84^o$ respectively. Integrating Eq. (\ref{Eq:b}) over 0.1 seconds, the sound image at $\phi = -54.8^o$ will be about 71.0 times more faint than the sonic boom, while the image at $\phi=2.84^o$ will be about 13.6 times more faint than the boom. As these times, angles, and brightnesses are within the limits of the modern technology of directional microphones, specifically microphones deploying a concave spherical or parabolic mirror \citep{1976ASAJ...59.1268C}. Therefore the existence of these sound-images -- and hence the pair-event nature of sonic booms -- should be falsifiable. 

\subsection{Cherenkov Light}

Another example where pair events are prominent is the initial detection of Cherenkov light. Although nothing can move faster than light in a vacuum, it is possible for an object moving at $c$ in vacuum to enter a medium with an index of refraction $n$ and hence move faster than $c_n = c/n$, the speed of light in that medium, at least initially. The resulting light emitted is well known as Cherenkov radiation \citep{1937PhRv...52..378C}. Cherenkov radiation is typically created by molecules in the medium itself being excited by the passing charged particle with the resulting de-excitation radiation emitted isotropically in the frame of the media \citep{1975clel.book.....J}. Here the molecules in the media are analogous to the static light bulbs in the previous discussions. Although the charged particle will decelerate after entering the medium, the magnitude of this deceleration will here be considered small.  

Cherenkov light is frequently described as being emitted into an annular cone with constant angular radius 
\begin{equation} \label{ThetaCherenkov}
\cos \theta = \frac{c}{n v} ,
\end{equation}
where $c$ is the speed of light in vacuum, and $v$ is the speed of the particle in the medium \citep{1975clel.book.....J}. Note that $v > c_n$ so that $\cos \theta < 1$ as expected. The angular vertex of the Cherenkov cone exactly coincides with the direction from which observers would see a pair event from the superluminal particle. When this angular cone intersects a plane, a ring of illumination occurs that is called a Cherenkov ring. Therefore, this ring is caused by a pair event. 

To the best of the author's knowledge, the dual nature of Cherenkov pair events has never yet been seen experimentally. A primary reason is because the identification of the perceived maximum in Cherenkov light intensity as pair events is relatively unknown. However, as with sonic booms, a simple verification experiment might confirm this. 

As an example, assume a highly energetic charged particle traveling near $c$ in air then enters a vat of water with $n=1.33$ and hence $c_n = c/1.33$. Further assume that this particle is so energetic that it does not slow significantly as it moves through the water, at least initially. Additionally, assume that the particle's Cherenkov light is incident on a high speed video camera, and that the minimum distance that the particle passes the camera is with $D_{min} = 1$ meter, for which, by definition $\phi = 0$. The first light seen by the camera will be a flash of Cherenkov light from the direction given by Eq. (\ref{Eq:PhiPair}) of $\phi_{pair} = -48.75^o$. Following Eq. (6) in \citet{2015PASA...32....1N}, one can find the times and positions of each subsequent Cherenkov light image. 

Isolating one wide-angle snapshot taken specifically at $10^{-9}$ seconds after the pair event and lasting $10^{-10}$ seconds, two separate images of the superluminal projectile will reach the camera: from directions $\phi \sim -72.96^o$ and $\phi=-3.21^o$ degrees respectively. Integrating Eq. (\ref{Eq:b}) over $10^{-10}$ seconds, the Cherenkov image at $\phi=72.96^o$ is about 210 times more faint that the main pair event, while the image at $\phi=-3.21^o$ is about 27.7 times more faint than the pair event. Recently, video cameras with frame rates below $10^{-9}$ sec have become prevalent \citep{2016SciA....2E1691C}. It therefore seems possible that modern technology can falsify the inherently pair-event nature of Cherenkov emission.

\section{Discussion and Conclusions}

Pair events and the subsequent separation of pair images are fundamental attributes of the detection of relative superluminal motion -- yet they are practically unknown. This work has predicted their existence for common physical phenomena involving massive objects in uniform linear motion and given examples of how they might be recovered in practice. 

Coherence effects involving the potential interference between the two images have been ignored in this work. Such effects may exist when two images occur within the coherence area of the source \citep{1985stop.book.....G}. Destructive interference could cause the two superluminal images to disappear, although this would have to occur just after a pair creation event or just before a pair annihilation event, as then both images could be within the coherence area of the source and of comparable brightness. In general, coherence effects are considered outside the scope of the present work, and may be studied in future works. 

If recovered experimentally, pair images might turn out to be more than just a new test of an unusual facet of perceived superluminal motion -- this recovery might uncover useful information. For example, assume that both images of a pair event are not only angularly resolved but photometrically measured after the pair event. Image angular locations and brightnesses could completely determine the kinematics of the projectile. One result could be the determination of the distance of the pair event. 

A complete kinematic solution would also, in turn, allow the observer to determine which image is moving time-forward and along the same direction as the projectile, and which image is time-reversed and moving in the opposite direction. It should then be possible to determine the past trajectory of the projectile by following the perceived future trajectory of this time-reversed image. Such ``backtracking" could be quite useful for determining, beyond the linear approximation, the history or origin of the superluminal projectile.

\section*{Acknowledgements}

I thank all of the people who have helped me to think carefully about the topics in this paper, in particular Qi Zhong and Ezequiel Medici. I also thank Jacek Borysow for a careful reading of the manuscript, and an anonymous referee for bringing up issues of coherence. 

\bibliography{PairEvents_PRA}

\end{document}